%% file: Qlearning_OF.tex
\documentclass[12pt, anon]{l4dc2024}
\usepackage{xcolor}
\usepackage{mathtools}
\DeclarePairedDelimiter{\norm}{\lVert}{\rVert}
\title[]{An efficient data-based off-policy Q-learning algorithm for optimal output feedback control of linear systems}
\usepackage{times, mathrsfs, mathtools}
\newcommand\undermat[2]{%
	\makebox[0pt][l]{$\smash{\underbrace{\phantom{%
					\begin{matrix}#2\end{matrix}}}_{\text{$#1$}}}$}#2}

\author{%
 \Name{Mohammad Alsalti} \Email{alsalti@irt.uni-hannover.de}\\
 \Name{Victor G. Lopez} \Email{lopez@irt.uni-hannover.de}\\
 \Name{Matthias A. M\"uller} \Email{mueller@irt.uni-hannover.de}\\
 \addr Leibniz University Hannover, Institute of Automatic Control, Hannover 30167, Germany%
}

\jmlrvolume{242}

\begin{document}

\maketitle

\begin{abstract}%
In this paper, we present a Q-learning algorithm to solve the optimal output regulation problem for discrete-time LTI systems. This off-policy algorithm only relies on using persistently exciting input-output data, measured offline. No model knowledge or state measurements are needed and the obtained optimal policy only uses past input-output information. Moreover, our formulation of the proposed algorithm renders it computationally efficient. We provide conditions that guarantee the convergence of the algorithm to the optimal solution. Finally, the performance of our method is compared to existing algorithms in the literature.
\end{abstract}


\input{sections/intro.tex}
\input{sections/main1.tex}
\input{sections/main2.tex}
\input{sections/sim.tex}
\input{sections/conc.tex}

\newpage
\acks{This work has received funding from the European Research Council (ERC) under the European Union’s Horizon 2020 research and innovation programme (grant agreement No 948679).}
\bibliography{references}

\end{document}

%% file: sections/intro.tex
\section{Introduction}\label{sec:intro}
Reinforcement learning (RL) \citep{Sutton18} is a suitable framework for learning optimal control policies for unknown dynamical systems \citep{Bertsekas12}. One of the most investigated problems in this area is the problem of optimal control of discrete-time linear time-invariant systems (LTI), also referred to as linear quadratic regulator (LQR) problem. For this problem, a control policy is sought such that the (weighted) deviation of the states from the origin and the control expenditure are minimized. Among many works that have addressed this problem in the RL literature are, e.g., \citep{Bradtke94, Kiumarsi17, Fazel18, Lopez21}.

When full state information is not available, the problem of optimal output regulation seeks to find a control policy such that the deviation of the outputs from the origin and the control expenditure are minimized. This problem has been addressed using RL approaches in, e.g., \citep{Lewis10, Kiumarsi15, Rizvi19}. These works present \textit{on-policy} algorithms, i.e., they require the application of each new policy to the system at each iteration. However, it is known that such algorithms lead, in general, to biased solutions due to the introduction of probing signals at each iteration, see, e.g., \citep{Lewis10, Kiumarsi15}.

Addressing optimal output regulation using \textit{off-policy} algorithms was done in \citep{Jiang20}, but this work required state measurements for the learning process. In \citep{Zhang23}, an off-policy algorithm for solving the $H_{\infty}$ optimal output regulation problem was presented. The proposed method requires additional constraints on the class of systems under consideration, specifically that the system is stabilizable by static output-feedback (see, e.g., \cite{kuvcera1995}), thus restricting its applicability. To solve the Bellman equation, both works mentioned above propose to use the batch least squares method, which becomes computationally inefficient for systems with high orders and large number of inputs. Finally, in those works it is unclear how to excite the system such that persistence of excitation conditions are met, or how to choose an initial stabilizing policy.

Apart from RL, other approaches have been recently proposed to solve the data-based optimal output regulation problem based on semi-definite programming (SDP) (cf. \cite{Dai23}), which make use of data-based representations of LTI systems (see, e.g., \cite{Willems05}). However, such SDPs are computationally expensive for large system dimensions.

\textbf{Contribution:} In this work, we present an efficient \textit{off-policy} Q-learning algorithm to solve the optimal output regulation problem. The proposed algorithm does not use model knowledge or state information and does not require that the system is stabilizable by static output-feedback. Instead, we use input-output data that are collected offline only once from the system. The resulting optimal policy is an output-feedback controller which uses only past input-output information. Our formulation relies on a suitable definition of a non-minimal state of the system which, along with an easily enforced persistence of excitation condition, renders the algorithm to be highly computationally efficient. We further show that the proposed algorithm is guaranteed to converge to the optimal gain at a quadratic rate of convergence, and a data-based method to design an initial stabilizing policy is given. This work serves as an extension of the algorithm presented in \citep{Lopez21} for the LQR problem with state information.

\textbf{Paper structure:} Section~\ref{sec:prob_form} formulates the problem and recalls existing results. Section~\ref{sec:main} includes the main contribution of this paper. Section~\ref{sec:examples} contains numerical examples and a comparison to existing methods. Finally, Section~\ref{sec:conclusion} discusses the results and concludes the paper.

\textbf{Notation:} We use $I_m$ to denote an $m\times m$ identity matrix and $0$ to denote a zero matrix of appropriate dimensions. A positive (semi-)definite matrix is denoted by $M\succ0$ ($M\succeq0$). For a sequence $\{s_k\}_{k=0}^{N-1}$ with $s_k\in\mathbb{R}^\eta$, we denote its stacked vector as $s = \begingroup\setlength\arraycolsep{2pt}\begin{bmatrix}s_0^\top &s_1^\top & \dots & s_{N-1}^\top\end{bmatrix}\endgroup^\top$ and a stacked window of it as $s_{[l,j]} = \begingroup\setlength\arraycolsep{2pt}\begin{bmatrix}s_l^\top &s_{l+1}^\top & \dots & s_{j}^\top\end{bmatrix}\endgroup^\top$ for $0\leq l<j$. The following definition of persistence of excitation of a sequence is used throughout the paper.
\begin{definition}[\cite{Willems05}]\label{def_PE}
	A sequence $s$ is said to be persistently exciting (PE) of order \(L\) if \(\textup{rank}(\mathscr{H}_{L}(s))=\eta L\), where the Hankel matrix $\mathscr{H}_L(s)$ takes the form
	\begin{equation}
		\mathscr{H}_L(s) = \begin{bmatrix}	s_{[0,L-1]} & s_{[1,L]} & \cdots & s_{[T-L,T-1]}	\end{bmatrix}
	\end{equation}
\end{definition}

In this work, we consider discrete-time LTI systems of the following form
\begin{equation}
	x_{k+1} = Ax_k + Bu_k,\qquad y_k = Cx_k,\label{eqn_LTI}
\end{equation}
where $x_k\in\mathbb{R}^n,\,u_k\in\mathbb{R}^m,\,y_k\in\mathbb{R}^p$ are the state, input and output vectors. The lag $\ell$ of the system (observability index) is defined as the smallest integer $j$ such that the matrix $\mathscr{O}_j=\begingroup\setlength\arraycolsep{3pt}\begin{bmatrix}
	C^\top & (CA)^\top & \cdots & (CA^{j-1})^\top
\end{bmatrix}^\top\endgroup$ has rank $n$. Note that the following relationship holds $\ell\leq n\leq p\ell$ \citep{Willems86}. The matrices $(A,B,C)$ are unknown but assumed to correspond to some minimal state-space realization and, hence, the pair $(A,B)$ is controllable and the pair $(A,C)$ is observable.

%% file: sections/main1.tex
\section{Problem formulation}\label{sec:prob_form}
We consider the optimal output regulation problem of the form
\begin{equation}
	\min\limits_{\{u_0,u_1,\cdots\}} \sum\limits_{k=0}^{\infty} c(y_k,u_k) = \min\limits_{\{u_0,u_1,\cdots\}}\sum\limits_{k=0}^{\infty} y_k^\top Q_y y_k + u_k^\top R u_k,\label{eqn_LQRy}
\end{equation}
with $Q_y,R\succ0$. Notice that the stage cost can also be written in terms of the state $x_k$ using the output equation in \eqref{eqn_LTI}. This results in a standard LQR problem with $c(y_k,u_k) = x_k^\top Q_x x_k + u_k^\top R u_k$, where $Q_x\coloneqq C^\top Q_y C\succeq0$ is, in general, a positive semi-definite matrix with $\sqrt{Q_x} = \sqrt{Q_y} C$. Since the pair $(A,C)$ is observable and since $Q_y\succ0$, the pair $(A,\sqrt{Q_x})$ is also observable. This, along with controllability of $(A,B)$, implies the existence and uniqueness of a solution to the LQR problem of the form $u_k=-K_x^*x_k$, where
\begin{equation}
	K_x^* = (R+B^\top P B)^{-1} B^\top P A,\label{eqn_KxModelBased}
\end{equation}
and $P\succ0$ is the unique solution of the discrete algebraic Riccati equation \citep{lewis2012optimal}. Since the matrices $(A,B,C)$ are unknown, one cannot evaluate the expression in \eqref{eqn_KxModelBased} and, moreover, since state measurements are not available, one cannot implement the policy $u_k=-K_x^*x_k$.

In \citep{Alsalti23b}, it was shown that a non-minimal state $z_k\in\mathbb{R}^{m\ell+n}$ for system \eqref{eqn_LTI} can be suitably constructed from past inputs and outputs\footnote{Although other non-minimal state definitions can be made (cf. \cite[Section 3.4]{Goodwin14}), the non-minimal state proposed in \citep{Alsalti23b} ensures the satisfaction of certain rank conditions which facilitate data-driven design of stabilizing controllers.}. Such a non-minimal state takes the form
\begin{equation}
	z_k = \begin{bmatrix}
		u_{[k-\ell,k-1]}\\ \Gamma\, y_{[k-\ell,k-1]}
	\end{bmatrix}\in\mathbb{R}^{m\ell+n},\label{eqn_state_z}
\end{equation}
where $\Gamma\in\mathbb{R}^{n\times p\ell}$ is a matrix that is computed from data as summarized in the following remark.
\begin{remark}[\cite{Alsalti23b}]\label{remark_computeGamma}
	Let $\{u_k,y_k\}_{k=-\ell}^{N-1}$ be input-output data collected from \eqref{eqn_LTI} with the input being PE of order $\ell+n+1$. Arrange the data in a Hankel matrix of the form $\begin{bsmallmatrix}
		\mathscr{H}_{\ell}(u_{[-\ell,N-2]})\\ \mathscr{H}_{\ell}(y_{[-\ell,N-2]})
	\end{bsmallmatrix}$ which, by minimality of the system and PE of the input, is guaranteed to have rank $m\ell+n$ (cf. \cite{Willems05}). One can now select $m\ell+n$ linearly independent rows of this matrix and arrange them in a matrix $Z_0\in\mathbb{R}^{m\ell+n\times N}$, while the remaining rows are arranged in $\Phi_0\in\mathbb{R}^{p\ell-n\times N}$. Next, one can find a permutation matrix $\Pi\in\mathbb{R}^{p\ell\times p\ell}$ which makes the following equation hold
	\begin{equation}
		\begin{bmatrix}
			\mathscr{H}_{\ell}(u_{[-\ell,N-2]})\\ \mathscr{H}_{\ell}(y_{[-\ell,N-2]})
		\end{bmatrix} = \begin{bmatrix}
			I_{m\ell} & 0\\ 0 & \Pi
		\end{bmatrix}\begin{bmatrix}
			Z_0\\ \Phi_0
		\end{bmatrix}.\label{eqn_separation_LIrows}
	\end{equation}
	Finally, the matrix $\Gamma\in\mathbb{R}^{n\times p\ell}$ can be found as follows $\Pi^{-1} = \begin{bmatrix}
		\Gamma^\top & G^\top
	\end{bmatrix}^\top$, where $G\in\mathbb{R}^{p\ell-n\times p\ell}$.
\end{remark}

The minimal state $x$ of the system in \eqref{eqn_LTI} and the non-minimal state $z$ in~\eqref{eqn_state_z} are related by $x_k = Tz_k$, where $T$ is a full row rank matrix (see \cite[Thm. 4]{Alsalti23b}) that depends on the unknown matrices $(A,B,C)$. Using $x_k = Tz_k$, one can express the optimal solution to \eqref{eqn_LQRy} as
\begin{equation}
	u_k = -K_x^*x_k = -K_x^*T z_k \eqqcolon -K_z^*z_k,\label{eqn_twopolicies}
\end{equation}
which is a control law that can be implemented, provided that the value of $K_z^*\in\mathbb{R}^{m\times m\ell+n}$ is found, since $z_k$ only consists of past inputs and outputs (cf. \eqref{eqn_state_z}). Notice that, given a non-minimal state $z_k$, the matrix $K_z^*$ is unique.

In \citep{Lopez21}, an efficient off-policy Q-learning algorithm was proposed to solve the LQR problem, assuming availability of state measurements. In this paper, we only consider availability of input-output data and, hence, the results of \citep{Lopez21} are not directly applicable. The goal of this work is to extend the results of \citep{Lopez21} to the case of output measurements and provide an efficient off-policy algorithm that converges to $K_z^*$ in \eqref{eqn_twopolicies} (thus solving \eqref{eqn_LQRy}), while maintaining the same uniqueness and convergence properties of \citep{Lopez21}. 

Before reviewing the results of \citep{Lopez21} in Section~\ref{sec:LQR_recap}, we first recall the following lemma, which provides conditions on the input such that certain rank conditions on the matrix of input and non-minimal state data are satisfied. This lemma is important for the results of Section~\ref{sec:main}.

\begin{lemma}[\cite{Alsalti23b}]\label{lemma_rankU0Z0}
	Let $\{u_k,y_k\}_{k=-\ell}^{N-1}$ be input-output data collected from system \eqref{eqn_LTI} with the input being PE of order $\ell+n+1$. Then $\textup{rank}\left(\begin{bsmallmatrix}
		Z_0\\ U_0
	\end{bsmallmatrix}\right) = m(\ell+1)+n,$ where
	\begin{equation}
		\begin{bmatrix}
			Z_0\\ U_0
		\end{bmatrix} = \begingroup\setlength\arraycolsep{3pt}\begin{bmatrix}
			z_0 & z_1 & \cdots & z_{N-1}\\
			u_0 & u_1 & \cdots & u_{N-1}
		\end{bmatrix}\endgroup \stackrel{\eqref{eqn_state_z}}{=} \begingroup\setlength\arraycolsep{3pt}\begin{bmatrix}
			I_{m\ell} & 0 & 0 \\ 0 & \Gamma & 0\\0 & 0 & I_m
		\end{bmatrix}\endgroup\begingroup\setlength\arraycolsep{3pt}\begin{bmatrix}
			u_{[-\ell,-1]} & u_{[-\ell+1,0]} & \cdots & u_{[N-\ell-1,N-2]}\\
			y_{[-\ell,-1]} & y_{[-\ell+1,0]} & \cdots & y_{[N-\ell-1,N-2]}\\ \hline
			u_0 & u_1 & \cdots & u_{N-1}
		\end{bmatrix}\endgroup.\label{eqn_defU0Z0}
	\end{equation}
\end{lemma}

\subsection{Q-learning algorithm for the LQR problem}\label{sec:LQR_recap}
In \citep{Lopez21}, an efficient off-policy Q-learning algorithm was proposed to solve the LQR problem, assuming availability of state measurements.	At each iteration of the algorithm, the value function $V^{(i)}(x_k)=x_k^\top P^{(i)} x_k$, for $P^{(i)}\succ0$, evaluates the cost of using a particular control input of the form $u = -K_x^{(i)} x$ from time $k$ onward. Similarly, a Q-function evaluates the cost of taking an arbitrary action at time $k$, then using the policy $u=-K_x^{(i)} x$ from time $k+1$ onward
\begin{equation}
		\mathcal{Q}^{(i+1)}(u_k,x_k) = \chi_k^\top \Theta_x^{(i+1)} \chi_k,\label{eqn_Qfcn}
\end{equation}
where $\chi_k\coloneqq\begin{bmatrix}x_k^\top & u_k^\top\end{bmatrix}^\top$ and
\begin{equation}
	\Theta_{x}^{(i+1)} = \begin{bmatrix}
		Q_x + A^\top P^{(i)} A & A^\top P^{(i)} B\\
		B^\top P^{(i)} A & R + B^\top P^{(i)} B
	\end{bmatrix} \eqqcolon \begin{bmatrix}
		\Theta_{xx}^{(i+1)} & (\Theta_{ux}^{(i+1)})^\top\\
		\Theta_{ux}^{(i+1)} & \Theta_{uu}^{(i+1)}
	\end{bmatrix}.\label{eqn_ThetaKx}
\end{equation}

The core of the proposed algorithm in \cite[Alg. 1]{Lopez21} consists of two parts: First, the solution of the Bellman equation
\begin{equation}
	\begin{aligned}
		\chi_k^\top \Theta_{x}^{(i+1)} \chi_k &= \chi_k^\top \overline{Q} \chi_k  + \begin{bmatrix}
			x_{k+1}^\top & (-K_x^{(i)} x_{k+1})^\top
		\end{bmatrix}^\top \Theta_{x}^{(i+1)} \begin{bmatrix}
			x_{k+1} \\ -K_x^{(i)} x_{k+1}
		\end{bmatrix},
	\end{aligned}\label{eqn_recursionKx1}
\end{equation}
where $\overline{Q} \coloneqq \textup{diag}(Q_x,R)$ and second, the policy improvement step. Specifically, an improved policy $K_x^{(i+1)}$ at the next iteration (in the sense that $V^{(i+1)}(x_k)\leq V^{(i)}(x_k)$) can be obtained using
\begin{equation}
	K_x^{(i+1)} = (\Theta_{uu}^{(i+1)})^{-1}\Theta_{ux}^{(i+1)}.\label{eqn_barKx}
\end{equation}

Equations \eqref{eqn_recursionKx1} and \eqref{eqn_barKx} were used to construct an iterative off-policy algorithm that converges to the solution \eqref{eqn_KxModelBased} of the LQR problem. Due to space constraints, we refer the reader to \citep{Lopez21} for more details on the construction of this algorithm. There, it was shown that solving the Bellman equation \eqref{eqn_recursionKx1} is equivalent to solving a discrete-time Lyapunov equation of the form
\begin{equation}
	\Theta_x^{(i+1)} = \overline{Q} + \Phi_i^\top \Theta_x^{(i+1)} \Phi_i,\qquad \textup{where }\, \Phi_i\coloneqq\begin{bmatrix}
		A & & B\\ -K_x^{(i)}A & & -K_x^{(i)}B
	\end{bmatrix}. \label{eqn_dlyapx}
\end{equation}

In \citep{Lopez21}, only positive definite matrices $\overline{Q}$ were considered. Clearly, if $\Phi_i$ is Schur stable, then for any $\overline{Q}\succ0$ there exists a unique and positive definite solution $\Theta_{x}^{(i+1)}\succ0$ to \eqref{eqn_dlyapx}. In the following lemma, the case where $\overline{Q}\succeq0$ is studied.
\begin{lemma}\label{lemma_dlyapx}
	If $K_x^{(i)}$ is a stabilizing gain, then for $\overline{Q}=\textup{diag}(Q_x,R)\succeq0$ (with $Q_x=C^\top Q_yC$ and $Q_y\succ0$) there exists a unique and positive definite solution $\Theta_{x}^{(i+1)}\succ0$ to \eqref{eqn_dlyapx}.
\end{lemma}
\begin{proof}
	Equation \eqref{eqn_dlyapx} is a discrete-time Lyapunov equation that is known to have a unique and positive definite solution $\Theta^{(i+1)}\succ0$ for any $\overline{Q}\succeq0$ if (i) $\Phi_i$ is Schur stable and (ii) $(\Phi_i,\sqrt{\overline{Q}})$ is observable \citep[p. 37]{lewis2012optimal}. Since $K_x^{(i)}$ is stabilizing, then matrix $\Phi_i$ is Schur stable (cf. \cite[Lemma 2]{Lopez21}). To show observability of the pair $(\Phi_i,\sqrt{\overline{Q}})$, it suffices to show that the pair $\left(\Phi_i,\begin{bsmallmatrix}
		C & 0\\ 0 & \sqrt{R}
	\end{bsmallmatrix}\right)$ is observable. This is because $\sqrt{\overline{Q}}$ can be expressed as
	\begin{equation}
		\sqrt{\overline{Q}} = \begin{bsmallmatrix}
			\sqrt{Q}_x & 0\\ 0 & \sqrt{R}
		\end{bsmallmatrix} = \begin{bsmallmatrix}
			\sqrt{Q_y} C & 0\\ 0 & \sqrt{R}
		\end{bsmallmatrix} = \begin{bsmallmatrix}
				\sqrt{Q_y} & 0\\ 0 & I_m
		\end{bsmallmatrix}\begin{bsmallmatrix}
			C & 0\\ 0 & \sqrt{R}
		\end{bsmallmatrix}.
	\end{equation}
	Since the matrix diag$(\sqrt{Q_y}, I_m)$ is invertible, the rank of the observability matrix of the pair $(\Phi_i,\sqrt{\overline{Q}})$ is equal to the rank of the observability matrix of $\left(\Phi_i,\begin{bsmallmatrix}
		C & 0\\ 0 & \sqrt{R}
	\end{bsmallmatrix}\right)$. To show observability of the latter, first recall that observability is preserved under invertible coordinate transformations. Now consider a transformation $T_{\phi}=\begin{bsmallmatrix}
		I_n & 0\\ -K_x^{(i)} & I_m
	\end{bsmallmatrix}$ such that $T_{\phi}^{-1}\Phi_iT_{\phi} = \begin{bsmallmatrix}
	A-BK_x^{(i)} & B\\ 0 & 0
\end{bsmallmatrix}$ and $\begin{bsmallmatrix}
C & 0\\ 0 & \sqrt{R}
\end{bsmallmatrix}T_{\phi} = \begin{bsmallmatrix}
C & 0\\ -\sqrt{R} K_x^{(i)} & \sqrt{R}
\end{bsmallmatrix}$. By the Popov–Belevitch–Hautus rank test (see \cite{Kailath80}), the pair $\left(\begin{bsmallmatrix}
A-BK_x^{(i)} & B\\ 0 & 0
\end{bsmallmatrix},\begin{bsmallmatrix}C & 0\\ -\sqrt{R} K_x^{(i)} & \sqrt{R}\end{bsmallmatrix}\right)$ is observable if and only if
	\begin{equation}
		\textup{rank}\left(\Omega_{\lambda}\right) = n+m,\quad \forall\lambda\in\mathbb{C},\qquad\quad\textup{where }\Omega_{\lambda}\coloneqq\begin{bmatrix}
			\lambda I - A + BK_x^{(i)} & \, & -B\\
			0 & \, & \lambda I\\
			C & \, & 0\\
			-\sqrt{R} K_x^{(i)} & \, & \sqrt{R}
		\end{bmatrix}.\label{eqn_PBHtest}
	\end{equation}
	For all $\lambda\neq0$, \eqref{eqn_PBHtest} holds since the pair  $\left(A-BK_x^{(i)},\begin{bsmallmatrix}C\\ -\sqrt{R} K_x^{(i)}\end{bsmallmatrix}\right)$ is observable (for any $K_x^{(i)}$) which follows from observability of the pair $(A,C)$ \citep[p. 72]{lewis2012optimal}. For $\lambda=0$, suppose for contradiction that rank$(\Omega_0)<n+m$. Then, there exists $r=\begin{bmatrix} r_1^\top & r_2^\top \end{bmatrix}^\top\neq0$, with $r_1\in\mathbb{R}^n$ and $r_2\in\mathbb{R}^{m}$, such that $\Omega_{0} r = 0$. This implies that $Cr_1=0$, $K_x^{(i)}r_1=r_2$ and that $Ar_1 = BK_x^{(i)}r_1-Br_2$. The last two equalities together imply that $Ar_1=0$, which together with $Cr_1=0$ and the observability of the pair $(A,C)$, imply that $r_1=0$. Finally, since $K_x^{(i)}r_1=r_2$ it follows that $r_2=0$, which contradicts the assumption that $r\neq0$. Therefore, \eqref{eqn_PBHtest} holds and the pair $\left(\begin{bsmallmatrix}
		A-BK_x^{(i)} & B\\ 0 & 0
	\end{bsmallmatrix},\begin{bsmallmatrix}C & 0\\ -\sqrt{R} K_x^{(i)} & \sqrt{R}\end{bsmallmatrix}\right)$ is observable.
\end{proof}
In the following section, we extend the results of \citep{Lopez21} to the case of output measurements and provide an efficient off-policy Q-learning algorithm that converges to $K_z^*$ in \eqref{eqn_twopolicies} (thus solving \eqref{eqn_LQRy}), while maintaining the same uniqueness and convergence properties of \citep{Lopez21}, which is the main contribution of the paper. 

%% file: sections/main2.tex
\section{Q-learning algorithm for the optimal output regulation problem}\label{sec:main}
As discussed in Section~\ref{sec:prob_form}, the solution of the output regulation problem \eqref{eqn_LQRy} has an equivalent parameterization in terms of the non-minimal state $z$ (see \eqref{eqn_twopolicies}). In this section, we exploit the state transformation $x_k=Tz_k$ in order to develop an iterative algorithm which asymptotically converges to the optimal solution of \eqref{eqn_LQRy} using only input-output data. First, notice that the following holds
\begin{equation}
	\chi_k = \begin{bmatrix}
		x_k\\ u_k
	\end{bmatrix} = \begin{bmatrix}
		T & 0\\ 0 & I_m
\end{bmatrix}\begin{bmatrix}
z_k\\ u_k
\end{bmatrix} \eqqcolon \mathscr{T} \zeta_k,\label{eqn_chiTozeta}
\end{equation}
where $\zeta_k=[z_k^\top \,\, u_k^\top]^\top\in\mathbb{R}^{m(\ell+1)+n}$ and $\mathscr{T}\in\mathbb{R}^{n+m\times m(\ell+1)+n}$ is a full row rank matrix. This allows us to express the Q-function defined in \eqref{eqn_Qfcn} as follows
\begin{equation}
	\begin{aligned}
		\mathcal{Q}^{(i+1)}(u_k,x_k) &= \chi_k^\top \Theta_{x}^{(i+1)}\chi_k \stackrel{\eqref{eqn_chiTozeta}}{=} \zeta_k^\top \mathscr{T}^\top\Theta_{x}^{(i+1)} \mathscr{T} \zeta_k \eqqcolon \zeta_k^\top \Theta_{z}^{(i+1)} \zeta_k,
	\end{aligned}\label{eqn_Qfcnz}
\end{equation}
where
\begin{equation}
	\Theta_z^{(i+1)} = \mathscr{T}^\top \Theta_x^{(i+1)}\mathscr{T} = \begin{bmatrix}
		T^\top \Theta_{xx}^{(i+1)} T & T^\top (\Theta_{ux}^{(i+1)})^\top\\
		\Theta_{ux}^{(i+1)} T & \Theta_{uu}^{(i+1)}
	\end{bmatrix} \eqqcolon \begin{bmatrix}
		\Theta_{zz}^{(i+1)} & (\Theta_{uz}^{(i+1)})^\top\\
		\Theta_{uz}^{(i+1)} & \Theta_{uu}^{(i+1)}
	\end{bmatrix},\label{eqn_Thetaz}
\end{equation}
with $\Theta_{zz}\in\mathbb{R}^{m\ell+n\times m\ell+n},\Theta_{uz}\in\mathbb{R}^{m\times m\ell+n}$ and $\Theta_{uu}\in\mathbb{R}^{m\times m}$. Recall that $\Theta_x^{(i+1)}$ is unique and positive definite at each iteration (see Lemma~\ref{lemma_dlyapx}). In contrast, $\Theta_z^{(i+1)}$ is unique but only positive semi-definite. This is due to the definition of $\Theta_z^{(i+1)}$ where $\Theta_x^{(i+1)}$ is pre- and post-multiplied by a full column and a full row rank matrix ($\mathscr{T}^\top$ and $\mathscr{T}$, respectively). Taking a closer look at the submatrices in \eqref{eqn_ThetaKx} and \eqref{eqn_Thetaz}, particularly, $\Theta_{uu}^{(i+1)}$, we see that this block is identical in both $\Theta_z^{(i+1)}$ and $\Theta_x^{(i+1)}$, which makes the following policy update step well-defined
\begin{equation}
	\begin{aligned}
		\arg\min\limits_{u}\mathcal{Q}^{(i+1)}(u_k,x_k) &= \arg\min\limits_{u} \chi_k^\top \Theta_x^{(i+1)}\chi_k =-(\Theta_{uu}^{(i+1)})^{-1}\Theta_{ux}^{(i+1)} x_k \stackrel{\eqref{eqn_barKx}}{=} -K_x^{(i+1)}x_k\\
		&= \arg\min\limits_{u} \zeta_k^\top \Theta_z^{(i+1)} \zeta_k = -(\Theta_{uu}^{(i+1)})^{-1}\Theta_{uz}^{(i+1)} z_k \eqqcolon -K_z^{(i+1)}z_k.
	\end{aligned}\label{eqn_twopolicies_i}
\end{equation}

Notice that the updated policy at each iteration has two equivalent parameterizations in terms of the minimal state $x$ and the non-minimal state $z$. Specifically, it holds that $K_z^{(i+1)}=K_x^{(i+1)} T$. The algorithm proposed in \citep{Lopez21} repeatedly solves for $K_x^{(i+1)}$ and was shown to converge to the optimal policy, i.e., $\lim_{i\to\infty}K_x^{(i+1)}=K_x^{*}$. However, that algorithm cannot be used to obtain $K_z^{(i+1)}$ for two reasons: (i) solving a set of $n+m$ equations of the form \eqref{eqn_recursionKx1} for $\Theta_{x}^{(i+1)}$ (compare Section~\ref{sec:LQR_recap}) requires state measurements, which are not available in our setting, and (ii) one cannot evaluate $K_z^{(i+1)} = K_x^{(i+1)}T$ because $T$ is also unknown and depends on model parameters $(A,B,C)$. Therefore, in the following we explain how one can solve for $\Theta_z^{(i+1)}$ directly, without requiring model knowledge and using only input-output data.

Recall from Lemma~\ref{lemma_rankU0Z0} that if $\{u_k,y_k\}_{k=-\ell}^{N-1}$ are input-output data collected from system~\eqref{eqn_LTI}, with $u$ being persistently exciting of order $\ell+n+1$, then the matrix $[Z_0^\top \,\, U_0^\top]^\top$ (defined as in \eqref{eqn_defU0Z0}) has full row rank (specifically, its rank is equal to $\nu\coloneqq m(\ell+1)+n$). Therefore, there exist $\nu$ linearly independent columns of this matrix of the form $\zeta_{k_j} = [ z_{k_j}^\top \,\, u_{k_j}^\top ]^\top$, for $j\in\{1,\ldots,\nu\}$. These vectors can be arranged in a square non-singular matrix of the form
\begin{equation}
	Z \coloneqq \begin{bmatrix}
			\zeta_{k_1} & \cdots & \zeta_{k_\nu}
		\end{bmatrix} = \begin{bmatrix}
			z_{k_1} & \cdots & z_{k_\nu}\\
			u_{k_1} & \cdots & u_{k_\nu}
		\end{bmatrix}\in\mathbb{R}^{\nu\times \nu}.\label{eqn_Zmat}
\end{equation}
Pre- and post-multiplying \eqref{eqn_dlyapx} by $Z^\top\mathscr{T}^\top$ and $\mathscr{T}Z$, respectively, with $\mathscr{T}$ as in \eqref{eqn_chiTozeta} results in
\begin{equation}
	Z^\top \mathscr{T}^\top\Theta_{x}^{(i+1)} \mathscr{T} Z = Z^\top \mathscr{T}^\top\overline{Q}\mathscr{T}Z + Z^\top \mathscr{T}^\top\Phi_i^\top \Theta_x^{(i+1)} \Phi_i\mathscr{T}Z.\label{eqn_Sylvester1}
\end{equation}
Since the product $\mathscr{T}Z$ results in a full row rank matrix, then a unique solution $\Theta_x^{(i+1)}$ of \eqref{eqn_dlyapx} implies the same unique solution of \eqref{eqn_Sylvester1}. Notice that \eqref{eqn_Sylvester1} still depends on model parameters (through $\mathscr{T},\overline{Q}$ and $\Phi_i$). Our goal is to arrive at a model-free version of this equation. By recalling the definition of $\Theta_z^{(i+1)}$ in \eqref{eqn_Thetaz}, one can rewrite \eqref{eqn_Sylvester1} as
\begin{equation}
	Z^\top \Theta_{z}^{(i+1)} Z = Z^\top \mathscr{T}^\top\overline{Q}\mathscr{T}Z + Z^\top \mathscr{T}^\top\Phi_i^\top \Theta_x^{(i+1)} \Phi_i\mathscr{T}Z.\label{eqn_Sylvester2}
\end{equation}
Moreover, by closely examining the rightmost term on the right hand side, one can use $x_k=Tz_k$ as well as the state equation in \eqref{eqn_LTI} to rewrite $\Phi_i\mathscr{T}Z$ as follows
\begin{align}
	\Phi_i\mathscr{T}Z &= \begin{bmatrix}
		AT & B\\ -K_x^{(i)}AT & -K_x^{(i)}B
	\end{bmatrix}\begingroup\setlength\arraycolsep{2pt}\begin{bmatrix}
		z_{k_1} & \cdots & z_{k_\nu}\\
		u_{k_1} & \cdots & u_{k_\nu}
	\end{bmatrix}\endgroup=\begingroup\setlength\arraycolsep{2pt}\left[\begin{matrix}
		Tz_{k_1+1} & \cdots\\ -{K_x^{(i)}T}z_{k_1+1} & \cdots
	\end{matrix} \begin{array}{c}Tz_{k_\nu+1} \\ -\undermat{\stackrel{\eqref{eqn_twopolicies_i}}{=}K_z^{(i)}}{K_x^{(i)}T}z_{k_\nu+1}\end{array} \,\, \right]\endgroup\notag\\
	&= \begin{bmatrix}
		T & 0\\ 0 & I_m
	\end{bmatrix}\begin{bmatrix}
		z_{k_1+1} & \cdots & z_{k_\nu+1}\\ -K_z^{(i)}z_{k_1+1} & \cdots & -K_z^{(i)}z_{k_\nu+1}
	\end{bmatrix}\eqqcolon \mathscr{T}\Sigma_{i,k+1},\label{eqn_fromPhiTZ_to_TSigma}
\end{align}
where we have defined $\Sigma_{i,k+1}$ as
\begin{equation}
	\Sigma_{i,k+1} \coloneqq \begin{bmatrix}
		z_{k_1+1} & \cdots & z_{k_\nu+1}\\ -K_z^{(i)}z_{k_1+1} & \cdots & -K_z^{(i)}z_{k_\nu+1}
	\end{bmatrix}\in\mathbb{R}^{\nu\times \nu}.\label{eqn_SigmaMat}
\end{equation}
Plugging this back into \eqref{eqn_Sylvester2}, we obtain the following equation
\begin{equation}
	\begin{aligned}
		Z^\top \Theta_{z}^{(i+1)} Z &= Z^\top \mathscr{T}^\top\overline{Q}\mathscr{T}Z + \Sigma_{i,k+1}^\top \underbrace{\mathscr{T}^\top \Theta_x^{(i+1)} \mathscr{T}}_{\stackrel{\eqref{eqn_Thetaz}}{=}\Theta_z^{(i+1)}} \Sigma_{i,k+1}.\label{eqn_Sylvester3}
	\end{aligned}
\end{equation}
In its current form, however, \eqref{eqn_Sylvester3} still depends on model parameters through the constant term $Z^\top \mathscr{T}^\top\overline{Q}\mathscr{T}Z$. This term can be equivalently expressed as
\begin{equation}
	\begin{aligned}
		Z^\top \mathscr{T}^\top\overline{Q}\mathscr{T}Z &= \begin{bmatrix}
			z_{k_1}^\top & u_{k_1}^\top\\
			\vdots & \vdots\\
			z_{k_\nu}^\top & u_{k_\nu}^\top
		\end{bmatrix} \begin{bmatrix}
		T^\top & 0\\ 0 & I_m
	\end{bmatrix}\begin{bmatrix}
			C^\top Q_y C & 0\\ 0 & R
		\end{bmatrix}\begin{bmatrix}
		T & 0\\ 0 & I_m
	\end{bmatrix}\begin{bmatrix}
			z_{k_1} & \cdots & z_{k_\nu}\\
			u_{k_1} & \cdots & u_{k_\nu}
		\end{bmatrix}\\
		&= \begin{bmatrix}
			(CTz_{k_1})^\top & u_{k_1}^\top\\
			\vdots & \vdots\\
			(CTz_{k_\nu})^\top & u_{k_\nu}^\top
		\end{bmatrix} \begin{bmatrix}
			Q_y & 0\\ 0 & R
		\end{bmatrix}\begin{bmatrix}
		CTz_{k_1} & \cdots & CTz_{k_\nu}\\
		u_{k_1} & \cdots & u_{k_\nu}
	\end{bmatrix}.
	\end{aligned}\label{eqn_rewritingZTQTZ}
\end{equation}
Using $x_k=Tz_k$, one can express the output as $y_k = Cx_k = CTz_k$. Furthermore, we define
\begin{equation}
	W\coloneqq \begin{bmatrix}
		y_{k_1} & \cdots & y_{k_\nu}\\
		u_{k_1} & \cdots & u_{k_\nu}
	\end{bmatrix}\in\mathbb{R}^{m+p\times \nu}.\label{eqn_Wmat}
\end{equation}
This, together with \eqref{eqn_rewritingZTQTZ} allows us to express $Z^\top \mathscr{T}^\top\overline{Q}\mathscr{T}Z = W^\top \hat{Q} W$ where we have defined $\hat{Q}\coloneqq\textup{diag}(Q_y,R)\succ0$. Finally, plugging back into \eqref{eqn_Sylvester3} results in
\begin{equation}
	Z^\top \Theta_{z}^{(i+1)} Z = W^\top \hat{Q} W + \Sigma_{i,k+1}^\top\Theta_z^{(i+1)}\Sigma_{i,k+1},\label{eqn_Sylvester}
\end{equation}
which is a special case of the generalized Sylvester matrix equation (specifically, a generalized discrete-time Lyapunov equation), for which efficient algorithms to solve it are known to exist \citep{Sasaki20}. Furthermore, this equation is model-free and can be solved for $\Theta_z^{(i+1)}$ at each iteration without requiring model-knowledge and only using input-output data.

\begin{algorithm}[t]
	\label{alg_LQRz}
	\floatconts{alg_LQRz}
	{\caption{Off-Policy Q-Learning for optimal output regulation problem}}
	{
		\begin{itemize}\setlength\itemsep{-0.5ex}
			\item[1.] Collect $N+\ell$ samples of data $\{ u_k,y_k \}_{k=-\ell}^{N-1}$ (with $N\geq m(\ell+n+1) + n - 1$) by applying a PE input of order $\ell+n+1$ to the system (cf. Definition~\ref{def_PE}). Then, construct $\{z_k\}_{k=0}^{N-1}$ as in \eqref{eqn_state_z} (cf. Remark~\ref{remark_computeGamma} and \cite{Alsalti23b}).
			\item[2.] Select $\nu=m(\ell+1)+n$ linearly independent vectors of the form $\zeta_{k_j} = [z_{k_j}^\top \,\, u_{k_j}^\top]^\top,\,j\in\{1,\ldots,\nu\}$. Then, construct the matrices $Z,W$ as in \eqref{eqn_Zmat} and \eqref{eqn_Wmat}, respectively.
			\item[3.] Let $i=0$ and find an initial stabilizing feedback policy $K_z^{(0)}$ (cf. Remark~\ref{remark:iniKz}).
			\item[4.] Using the gain $K_z^{(i)}$, construct the matrix $\Sigma_{i,k+1}$ in \eqref{eqn_SigmaMat}.
			\item[5.] Solve the matrix equation \eqref{eqn_Sylvester} for $\Theta_z^{(i+1)}$ and update $K_z^{(i+1)}=(\Theta_{uu}^{(i+1)})^{-1} \Theta_{uz}^{(i+1)}$.
			\item[6.] If $\| K_z^{(i+1)} - K_z^{(i)} \| > \varepsilon$ for some $\varepsilon > 0$, let $i=i+1$ and go to Step 4. Otherwise, stop.
		\end{itemize}
	}
\end{algorithm}
We now introduce the main result of this paper, which is an iterative off-policy Q-learning algorithm that solves \eqref{eqn_LQRy}. This is summarized in Algorithm~\ref{alg_LQRz}, which takes as an input previously collected, persistently exciting, input-output data as well as an initial stabilizing policy (see Remark~\ref{remark:iniKz} below for a discussion). At each iteration, we use the collected data to solve \eqref{eqn_Sylvester} for $\Theta_z^{(i+1)}$ and obtain an improved policy $K_z^{(i+1)}=(\Theta_{uu}^{(i+1)})^{-1}\Theta_{uz}^{(i+1)}$ as in \eqref{eqn_twopolicies_i}. The policy can now be used to start a new iteration, by updating the matrix $\Sigma_{i,k+1}$ in \eqref{eqn_Sylvester} (see \eqref{eqn_SigmaMat}). The algorithm is terminated once the difference between two successive policies is below a user-defined threshold $\varepsilon>0$.

\begin{remark}\label{remark:iniKz}
	One can use \citep[Alg. 2]{Lopez21} with $\{u_k,z_k\}_{k=0}^{N-1}$ to obtain a deadbeat controller $u=-K_{\textup{db}}z$ which can be used as an initial stabilizing policy for Algorithm~\ref{alg_LQRz}. Alternatively, an initial stabilizing policy can be obtained by solving an LMI (cf. \cite[Th. 5]{Alsalti23b}).
\end{remark}	

The following theorem shows that Algorithm~\ref{alg_LQRz} enjoys all of the theoretical guarantees of its state-feedback counterpart \citep[Alg. 1]{Lopez21}.
\begin{theorem}
	Let the conditions in Lemma~\ref{lemma_rankU0Z0} hold. Given an initial stabilizing policy $K_z^{(0)}$, the solution $\Theta_z^{(i+1)}$ of \eqref{eqn_Sylvester} exists and is unique. Moreover, every policy $K_z^{(i+1)}$ obtained at each iteration of Algorithm~\ref{alg_LQRz} is stabilizing and $\lim_{i\to\infty}K_z^{(i)}=K_z^{*}$ converges quadratically.
\end{theorem}
\begin{proof}
	By construction, the matrix equations in \eqref{eqn_Sylvester} and \eqref{eqn_dlyapx} are equivalent in the sense that $\Theta_z^{(i+1)}=\mathscr{T}^\top\Theta_x^{(i+1)}\mathscr{T}$ (see the developments below \eqref{eqn_Zmat}). This implies that if $K_z^{(0)}=K_x^{(0)}T$ is used as an initial matrix in Algorithm~\ref{alg_LQRz}, and $K_x^{(0)}$ is used as an initial matrix in \cite[Alg. 1]{Lopez21}, then $K_z^{(i+1)}=K_x^{(i+1)}T$ at every iteration $i$, compare \eqref{eqn_twopolicies_i}. Existence and uniqueness of the solution $\Theta_x^{(i+1)}$ to \eqref{eqn_dlyapx} is given by Lemma~\ref{lemma_dlyapx}, which implies existence and uniqueness of $\Theta_z^{(i+1)}$. Moreover, since $u=-K_x^{(i+1)}x$ is stabilizing for system \eqref{eqn_LTI} and since $x=Tz$, it holds that $K_z^{(i+1)}=K_x^{(i+1)}T$ is also stabilizing. Finally, equivalence of \eqref{eqn_Sylvester} and \eqref{eqn_dlyapx} shows that Algorithm~\ref{alg_LQRz} and \cite[Alg. 1]{Lopez21} converge at the same quadratic rate.
\end{proof}
\begin{remark}
	In the presence of small magnitudes of measurement noise, \cite[Alg. 1]{Lopez21} converges to a solution that differs from the optimal one $K_x^*$ by an amount that is dependent on the noise bound. Although a comprehensive discussion about the effect of noise in the data is out of the scope of this paper, we expect that Algorithm~\ref{alg_LQRz} possesses similar inherent robustness properties.
\end{remark}

%% file: sections/sim.tex
\section{Comparisons and Simulation examples}\label{sec:examples}
In this section, we discuss the main differences of Algorithm~\ref{alg_LQRz} compared to existing algorithms that were proposed to solve problem \eqref{eqn_LQRy}. First, notice that \citep{Jiang20} uses \textit{state} measurements and thus it is not a purely input-output method. The work by \citep{Zhang23} restricts the class of systems \eqref{eqn_LTI} to those that are stabilizable by a static output feedback (cf. \cite{kuvcera1995}). In contrast, our method is applicable to any LTI system of the form~\eqref{eqn_LTI}. Moreover, both works \citep{Jiang20, Zhang23} propose to solve the Bellman equation using the batch least squares approach, and use a number of data points that scales with $n^2+m^2+nm$. In contrast, we only require $N+\ell$ data points (where $N\geq  m(\ell+n+1) + n - 1$) and solve the Bellman equation using efficient solvers for the generalized discrete-time Lyapunov equation. Finally, unlike \citep{Jiang20, Zhang23}, we provide methods to guarantee persistence of excitation (cf. Lemma~\ref{lemma_rankU0Z0}) as well as the design of an initial stabilizing policy from data (cf. Remark~\ref{remark:iniKz}).

A different approach to solving the optimal output regulation problem was proposed in \citep{Dai23}, where a policy $u_k=\mathcal{K}\xi_k$, with $\xi_k=[u_{[k-n,k-1]}^\top \,\, y_{[k-n,k-1]}^\top]^\top$, can be found by solving a data-dependent SDP. However, such SDPs are computationally expensive for large system dimensions. For the state-feedback case, computational complexity of \citep[Alg. 1]{Lopez21} was studied and shown to outperform SDP-based methods that solve the LQR problem (e.g., \citep{Persis21}), see \citep[Sec. VII]{Lopez21}. We expect similar conclusions for Algorithm~\ref{alg_LQRz} in comparison to \citep{Dai23} for the output-feedback case.

In the following, we will compare the performance of our proposed Algorithm~\ref{alg_LQRz} to the SDP-based method proposed in \citep{Dai23} in a numerical case study. We use the definition of the non-minimal state $z_k$ in \eqref{eqn_state_z}, instead of $\xi_k$, when solving the SDP. This is because, unlike $\xi_k$, the state $z_k$ is such that $[Z_0^\top \,\, U_0^\top]^\top$ has full row rank (cf. Lemma~\ref{lemma_rankU0Z0}). We do not compare against any of the RL off-policy methods mentioned above because (i) \citep{Jiang20} requires state measurements, which we do not and (ii) \citep{Zhang23} imposes additional assumptions on the class of systems, thus rendering any comparison inconsistent.

For simulations, we consider random LTI systems (both open-loop stable and unstable) of different orders $n=\{3,5,10,30,50\}$. For each state dimension, we generate 100 systems with number of inputs and outputs shown in Table~\ref{table:comparison}. For each system, we run the Q-learning (QL) Algorithm~\ref{alg_LQRz} (for 10 iterations only, regardless of the system dimension) and the method proposed by \citep{Dai23} based on  an SDP (adapted as described above) in order to solve \eqref{eqn_LQRy} with $Q_y = 100I_p, R=I_m$. Since we use the non-minimal state $z_k$ in the SDP, the two approaches (QL and SDP) use the same number of data points $N+\ell$, which we set to the lower bound, i.e., $N+\ell= (m+1)(\ell+n+1) -2$. Hence, we use the same input-output data which we collect\footnote{For open-loop unstable systems, using one (long) experiment results in output trajectories with arbitrarily large magnitudes. Instead, we use multiple shorter experiments whose inputs are \textit{collectively PE}, see \citep{vanWaarde20}.} once offline by applying a PE input to the system sampled from a uniform random distribution $(\mathcal{U}(-1,1))^m$. For each run, we record the total run time as well as the error from the true optimal policy $K_z^*$ given by $\epsilon=\norm*{K_z^* - K_z^{\textup{sol}}}_2$, where $K_z^{\textup{sol}}$ is the solution returned by each scheme. Finally, we average the results over all 100 runs and report them in Table~\ref{table:comparison}. For all simulations, Matlab R2021a was used on an Intel i7-10875H (2.30 GHz) with 16 GB of memory. The Matlab source code is available at \url{https://doi.org/10.25835/zmlriehg}.

It can be seen that Algorithm~\ref{alg_LQRz} (QL) outperforms the SDP method from \citep{Dai23} in terms of both total run time and the average error $\epsilon$. For system dimensions $n=\{5,10\}$, the SDP method was observed to be (highly) numerically unstable and, in many instances, no solution was found (using CVX \citep{cvx} with three different solvers SDPT3, MOSEK and SeDuMi). For those instances where no solution was found, we discard the experiment and repeat it until 100 successful runs are completed\footnote{Out of these 100 successful runs of the SDP method, some did not return the optimal solution which explains the (large) average errors reported in Table~\ref{table:comparison}}. For even larger system dimensions, a solution to the SDP was never found on our machine for open-loop unstable systems. In comparison, our proposed Q-learning algorithm successfully converges to the optimal solution relatively fast, even for large system orders, due to efficiently solving the Bellman equation \eqref{eqn_Sylvester} using \texttt{dlyap} in Matlab. Note that QL requires an initial stabilizing policy, which was computed\footnote{The computation time of the initial stabilizing policy is included in the values reported in Table~\ref{table:comparison}.} as in \cite[Alg.~2]{Lopez21}, whereas the SDP method does not require any initialization.

\begin{table}[!t]
	\caption{Numerical simulations on 100 random systems of different dimensions.}
	\vspace{0.5em}
	\label{table:comparison}
	\centering
	\begin{tabular}{ | ccc | cc | cc | }
		\hline
		\multicolumn{3}{|c|}{Complexity} & \multicolumn{2}{c|}{QL (Algorithm~\ref{alg_LQRz})}& \multicolumn{2}{|c|}{SDP \citep{Dai23}}\\
		\hline
		$n$ & $p$ & $m$ & $\textup{avg run time } [\mathrm{s}]$ & $\epsilon$ & $\textup{avg run time } [\mathrm{s}]$ & $\epsilon$\\ \hline
		$3$ & $2$ & $1$ & $0.001$ & $5.55\times10^{-13}$ & $0.398$ & $9.04\times 10^{-6}$\\
		$5$ & $3$ & $2$ & $0.002$ & $1.60\times10^{-10}$ & $1.256$ & $0.3887$\\
		$10$ & $6$ & $5$ & $0.005$ & $6.47\times10^{-9}$ & $14.130$ & $22.5708$\\
		$30$ & $15$ & $10$ & $0.024$ & $2.37\times 10^{-7}$ & n/a & n/a\\
		$50$ & $20$ & $15$ & $0.116$ & $1.66\times 10^{-4}$ & n/a & n/a\\
		\hline
	\end{tabular}
\end{table}

%% file: sections/conc.tex
\section{Conclusion}\label{sec:conclusion}
In this paper, we presented a data-based off-policy Q-learning algorithm to solve the optimal output regulation problem for discrete-time LTI systems. This was done by extending an existing Q-learning algorithm for the LQR problem to the case of optimal output-feedback control, and exploiting the definition of a suitable non-minimal state of the system which depends only on past inputs and outputs. Our proposed algorithm only uses previously collected, persistently exciting input-output data. Persistence of excitation is easily enforced by a rank condition on the Hankel matrix of the input data. Furthermore, the algorithm uses an initial stabilizing policy which can be designed from data. Finally, it was shown that the algorithm asymptotically converges to the optimal policy. When compared to an existing method from the literature, our algorithm was shown to outperform it both in terms of total run time as well as scalability to large dimensional systems.